%% file: KQ_review.tex
\documentclass[11pt]{article}
\usepackage{graphicx} 
\usepackage{amsmath}
\usepackage{amsfonts}
\usepackage{arydshln}

\usepackage{wrapfig}
\usepackage{tikz-cd} 
\usepackage{tkz-graph}
\usepackage{quiver}

\usepackage{authblk} 
\usepackage{orcidlink}

\usepackage[a4paper,top=3.2cm,width=16cm, bottom=3.2cm, left=2cm]{geometry}
\usepackage{hyperref}
\hypersetup{
    colorlinks=true,
    linkcolor=blue,
    filecolor=magenta,      
    urlcolor=cyan,
    citecolor=blue
}
\newtheorem{thm}{Theorem}[subsection]

\newtheorem{coj}[thm]{Conjecture}

\newcommand{\overcross}{
 {\mathchoice
  {\includegraphics[height=1.6ex]{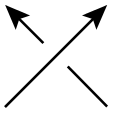}}
  {\includegraphics[height=1.6ex]{Figures/overcrossing.png}}
  {\includegraphics[height=1.2ex]{Figures/overcrossing.png}}
  {\includegraphics[height=0.9ex]{Figures/overcrossing.png}}
 }
}
\newcommand{\undercross}{
 {\mathchoice
  {\includegraphics[height=1.6ex]{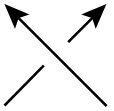}}
  {\includegraphics[height=1.6ex]{Figures/undercrossing.png}}
  {\includegraphics[height=1.2ex]{Figures/undercrossing.png}}
  {\includegraphics[height=0.9ex]{Figures/undercrossing.png}}
 }
}
\newcommand{\nocross}{
 {\mathchoice
  {\includegraphics[height=1.6ex]{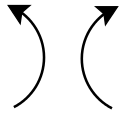}}
  {\includegraphics[height=1.6ex]{Figures/no_crossing.png}}
  {\includegraphics[height=1.2ex]{Figures/no_crossing.png}}
  {\includegraphics[height=0.9ex]{Figures/no_crossing.png}}
 }
}

\title{Knot-quiver correspondence: a brief review}

\author[1]{Piotr Kucharski \orcidlink{0000-0002-9599-5658}\thanks{piotr.kucharski@mimuw.edu.pl}}

\author[2]{\break Dmitry Noshchenko \orcidlink{0000-0002-9639-5603}\thanks{dsnoshchenko@stp.dias.ie}}

\affil[1]{Institute of Mathematics, University of Warsaw, ul. Banacha 2, 02-097 Warsaw, Poland}

\affil[2]{School of Theoretical Physics, Dublin Institute for Advanced Studies,\break 10 Burlington Road, Dublin 4, D04 C932, Ireland}

\date{April 2025}

\begin{document}

\maketitle

\begin{abstract}
\centering
    This note is an overview of the knot-quiver correspondence, which relates symmetric quivers and their partition functions, a.k.a. motivic Donaldson-Thomas generating series, to quantum invariants of knots and links in $S^3$.
\end{abstract}

\tableofcontents

\section{Introduction}

\subsection{Knots}\label{sec:knots}

\begin{figure}[h!]
    \centering
    \includegraphics[scale=0.35]{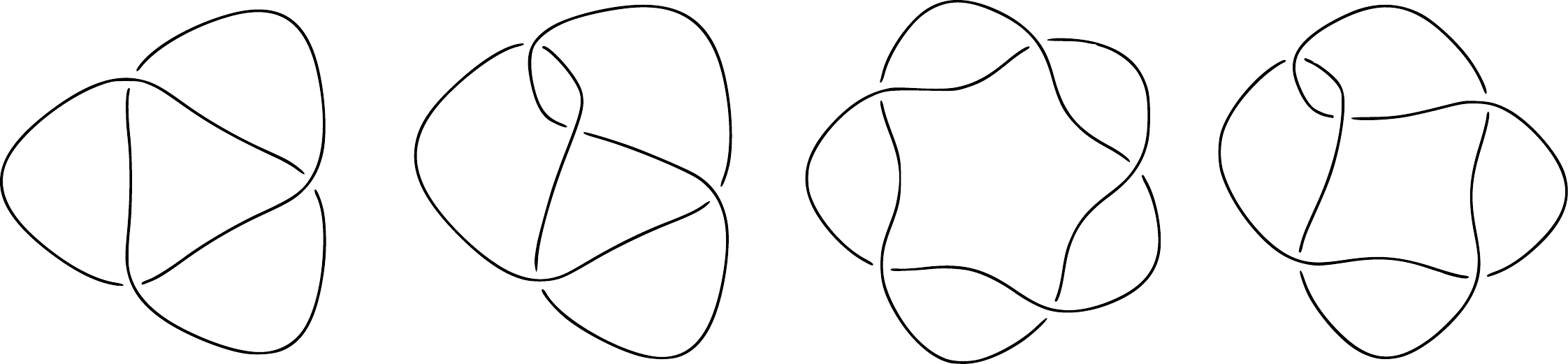}
    \caption{Knots $3_1$, $4_1$, $5_1$ and $5_2$ (generated by KnotScape)}
    \label{fig:few knots}
\end{figure}

In mathematics, knots are embeddings of a circle into 3-dimensional space which, intuitively speaking, means that the ends of the rope are glued after tying the knot. Among all knots, the prime ones are of the main interest -- by analogy with prime numbers, such knots cannot be presented as a connected sum of any other knots. Throughout this note, we will focus on the trivial knot (unknot) along with the first few prime knots, shown in Figure \ref{fig:few knots}.\footnote{For the reader unfamiliar with knot theory, we refer to a classic book by Dale Rolfsen \cite{Rolfsen}.}

The main goal of knot theory is understanding which knots are topologically identical, and which are different.
So far there is no perfect solution, but a very efficient one is to assign some mathematical objects to knots in a way that is topologically invariant. From the perspective of a relation to quivers, the most relevant invariant of a knot $K$ is the \textit{HOMFLY-PT polynomial}\footnote{This name is an abbreviation of the first letters of the last names of the authors of \cite{HOMFLY,PT}} $P_K(a,q)$ which can be defined using the skein relation
\begin{equation}\label{eq:homfly-pt skein}
    aP[\overcross]-a^{-1}P[\undercross]=(q-q^{-1})P[\nocross]\, .
\end{equation}
Note that it should be understood as a linear relation between polynomials corresponding to knots as well as links\footnote{Links are embeddings of several circles instead of only one.} which differ only by the fragment depicted in the equation above, plus the initial condition for the unknot: $P_{\bigcirc}(a,q)\equiv P[\bigcirc]=1$.

In the quest of finding more sophisticated structures among knot invariants, one can generalise HOMFLY-PT polynomials by introducing a whole infinite family of polynomials associated to a given knot, parametrised by an integer number $r$ called \emph{colour}. We denote such polynomials  $P_{K,r}(a,q)$  and call them \textit{coloured HOMFLY-PT polynomials}.
The notion of colour comes from the gauge-theoretic interpretation of these invariants, which will be briefly discussed below.
On the other hand, a simple diagrammatic interpretation of coloured invariant $P_{K,r}(a,q)$ is a consideration of a standard invariant $P_{K,1}(a,q)\equiv P_{K}(a,q)$, but corresponding to an $r$-cabling of $K$. Of course, if $r=1$ then the $r$-cabling of $K$ is the same as $K$ itself.
One can then apply the usual skein relation (\ref{eq:homfly-pt skein}) for any given number $r$ of cables, in order to compute these invariants. For example, all coloured invariants of an unknot are equal to 1 by definition. On the other hand, for the trefoil knot $3_1$ the first two such polynomials are:
\begin{align}
    P_{3_1,1}(a,q) = &\ -a^4+a^2 q^2+\frac{a^2}{q^2} \\
    P_{3_1,2}(a,q) = &\ a^8 q^6-a^6 q^8-a^6 q^6-a^6 q^2-a^6+a^4 q^8+a^4 q^4+\frac{a^4}{q^4}+a^4 q^2
\end{align}

It is convenient to combine $P_{K,r}(a,q)$ into the \textit{HOMFLY-PT generating series}:
\begin{equation}\label{eq:homfly gen series}
    P_K(x,a,q)=\sum_{r=0}^{\infty}\frac{P_{K,r}(a,q)}{(q^2;q^2)_r}x^r,
\end{equation}
where $x$ is a formal parameter and 
\begin{equation}
    (\alpha;q^{2})_{n} = \prod_{k=0}^{n-1}(1-\alpha q^{2k})
\end{equation}
is the~$q$-Pochhammer symbol. The choice of denominator in (\ref{eq:homfly gen series}) may seem a bit random at this moment, but will be made clear in the next section.

HOMFLY-PT invariants play special role in Chern-Simons gauge theory with gauge group $G=SU(N)$, as shown by Witten \cite{Wit87} based on earlier ideas of Atiyah and Schwartz. Namely, $P_{K,r}(a,q)$ can be understood as expectation value of a Wilson loop observable around a knot $K$, where the colour $r$ corresponds to a symmetric representation $S^r$ of the gauge group (one usually says that the knot is ``coloured'' by a representation). These ideas were further integrated into string theory picture by Gopakumar-Ooguri-Vafa \cite{GV9811,OV9912}, suggesting the duality between $SU(N)$ Chern-Simons theory and open topological strings. This led to the interpretation of generating series (\ref{eq:homfly gen series}) as a partition function
for open string amplitudes in a resolved conifold. 
We will not go into details here, but encourage an interested reader to learn more about knots and their relation to physics using e.g. \cite{MarCS,GukovSaberi}.

\subsection{Quivers}
Quivers are basically directed graphs, so they consist of nodes connected by arrows. It is very convenient to represent a quiver $Q$ in form of a matrix whose entries  $C_{ij}$ are equal to the number of arrows from node $i$ to node $j$.\footnote{Diagonal entries encode the number of loops.}  A~quiver is called symmetric if $C_{ij}=C_{ji}$. An example of symmetric quiver is shown below:
\[\begin{tikzcd}
	\bullet & \bullet & \begin{bmatrix}1 & 1 \\ 1 & 0\end{bmatrix}
	\arrow[from=1-1, to=1-1, loop, in=60, out=120, distance=5mm]
	\arrow[curve={height=-6pt}, from=1-1, to=1-2]
	\arrow[curve={height=-6pt}, from=1-2, to=1-1]
\end{tikzcd}\;.\]

Quivers are especially important in the context of representation theory. A quiver representation is a~map which assigns a vector space  $\mathbb{C}^{d_i}$ to node $i$ and a linear map $\mathbb{C}^{d_i}\to\mathbb{C}^{d_j}$ to arrow $i\to j$. Since the structure of possible representations crucially depends on the dimensions, it is convenient to assign them into a dimension vector  $\boldsymbol{d}=(d_1,\ldots,d_{m})$, where $m$ is the number of quiver nodes. 

Mathematicians are usually interested in the~topological properties of the space of all possible representations of a given quiver, which are encoded in the motivic Donaldson-Thomas invariants (DT invariants) $\Omega_{(d_1,\ldots,d_{m}),s}=\Omega_{\boldsymbol{d},s}$. Simplifying a little, DT invariants can be thought of as a version of Betti numbers for the space of all quiver representations. For a~symmetric quiver $Q$ there exist a surprising shortcut which allows us to access its DT invariants simply by analysing the following generating series:\footnote{In the literature also sometimes referred to as ``motivic generating series of a quiver $Q$''.}
\begin{equation}
    P_Q(\boldsymbol{x},q) =
    \sum_{\boldsymbol{d}\in \mathbb{N}^{m}}(-q)^{\boldsymbol{d} \cdot C\cdot\boldsymbol{d}}\frac{\boldsymbol{x}^{\boldsymbol{d}}}{(q^{2};q^{2})_{\boldsymbol{d}}}
    = \sum_{d_1,\dots,d_{m}\geq 0}(-q)^{\sum_{i,j=1}^{m} C_{ij}d_i d_j}\prod_{i=1}^{m}\frac{x_i^{d_i}}{(q^{2};q^{2})_{d_i}},    \label{PQ}
\end{equation}
where a~generating parameter $x_i$ is assigned to each vertex $i$. We call $P_Q(\boldsymbol{x},q)$ the \textit{quiver partition function} of $Q$ and the shortcut to DT invariants runs through the product decomposition. More precisely, it turns out that $ P_Q(\boldsymbol{x},q)$ can be rewritten as product of infinite versions of~$q$-Pochhammer symbols called quantum dilogarithms $(\alpha;q^{2})_{\infty} = \prod_{k=0}^{\infty}(1-\alpha q^{2k})$ and DT invariants are the exponents\footnote{The product over $\boldsymbol{d}\in \mathbb{N}^{m}\setminus \boldsymbol{0}$ means almost the same as over $d_1,\dots,d_{m}\geq 0$, we just exclude $d_1=\dots d_{m}=0$}:
\begin{equation}\label{PQ-product}
    P_Q(\boldsymbol{x},q) = \prod_{\boldsymbol{d}\in \mathbb{N}^{m}\setminus \boldsymbol{0}} \prod_{s\in\mathbb{Z}}(\boldsymbol{x}^{\boldsymbol{d}}q^s;q^2)_{\infty}^{\Omega_{\boldsymbol{d},s}} = \prod_{\boldsymbol{d}\in \mathbb{N}^{m}\setminus \boldsymbol{0}} \prod_{s\in\mathbb{Z}} \prod_{k\geq 0} \Big(1 - (x_1^{d_1}\cdots x_{m}^{d_{m}}) q^{2k+s} \Big)^{\Omega_{(d_1,\ldots,d_{m}),s}}.
\end{equation}
It is convenient to combine DT invariants in a~generating series
\begin{equation}
    \Omega(\boldsymbol{x},q) = \sum_{\boldsymbol{d}\in \mathbb{N}^{m}\setminus \boldsymbol{0}} \Omega_{\boldsymbol{d}}(q)\, \boldsymbol{x}^{\boldsymbol{d}} =  \sum_{\boldsymbol{d}\in \mathbb{N}^{m}\setminus \boldsymbol{0}}\sum_{s\in \mathbb{Z}} \,\Omega_{(d_1,\ldots,d_{m}),s}\, x_1^{d_1}\cdots x_{m}^{d_{m}} q^s.   \label{Omega-series}
\end{equation}

A non-trivial fact is that invariants $\Omega_{(d_1,\ldots,d_{m}),s}$ defined via the~decomposition (\ref{PQ-product}) are integer, and multiplied by $(-1)^{j+1}$ become positive \cite{KS1006,Efi12}.  More information about quivers and their representations can be found in \cite{Gei06,Rei11,Rei12}.

\section{Knot-quiver correspondence}

In this section we will discuss our central topic, which is the correspondence between symmetric quivers and knots.
It turned out to be a rather surprising relation -- before these findings no such explicit links between knot theory and quiver representation theory were known, as they are a priori two very different subjects. As we shall see, the main statement of the correspondence is simple, but rooted deeply in physics and geometry of knots.

\subsection{Equality of generating series}

Knot-quiver correspondence was initially stated as the equality between the generating series of coloured HOMFLY-PT polynomials of a knot and the partition function of a symmetric quiver \cite{KRSS1707short,KRSS1707long}:
\begin{equation}
    P_K(x,a,q) = \left. P_Q(\boldsymbol{x},q) \right|_{x_i=x a^{a_i} q^{q_i-C_{ii}}}\,.
\end{equation}
Parameters $a_i,q_i$ are the exponents of $a,q$ in the standard HOMFLY-PT polynomial $P_{K,1}(a,q)\equiv P_{K}(a,q)$, whereas $C_{ii}$ is the number of loops at node $i$. 

Let us see how the correspondence works on the example of the trefoil knot. Its HOMFLY-PT polynomials coloured by symmetric representations are given by
\begin{equation}
    P_{3_1,r}(a,q)=  \frac{a^{2r}}{q^{2r}}\sum_{k=0}^r q^{2k(r+1)}\frac{(q^2;q^2)_r \big(\frac{a^2}{q^2};q^2\big)_k}{(q^2;q^2)_{r-k} (q^2;q^2)_k}  .    \label{Pr-31-homfly}
\end{equation}
Using the identity \cite{KRSS1707long}
\begin{equation}
    (x;q)_n= \sum\limits_{\alpha+\beta=n} (-x)^{\alpha} q^{\frac{1}{2}\alpha(\alpha-1)}\frac{(q;q)_{n}}{(q;q)_{\alpha}(q;q)_{\beta}},  \label{qpoch-sum}
\end{equation}
we can rewrite \eqref{Pr-31-homfly} as
\begin{equation}
     P_{3_1,r}(a,q)=  \frac{a^{2r}}{q^{2r}}\sum_{k=0}^r q^{2k(r+1)}
    \sum_{i=0}^k \frac{(q^2;q^2)_r  \big( - \frac{a^2}{q^2} \big)^i q^{i(i-1)} }{(q^2;q^2)_{r-k} (q^2;q^2)_i (q^2;q^2)_{k-i}} ,
\end{equation}
so the HOMFLY-PT generating series takes the form
\begin{equation}
    P_{3_1}(x,a,q) = \sum_{r=0}^{\infty} \frac{ P_{3_1,r}(a,q)}{(q^2;q^2)_r} x^r =   \sum_{r=0}^{\infty} \sum_{k=0}^r 
    \sum_{i=0}^k (-q)^{2kr+i^2}\frac{x^r a^{2r+2i}q^{-2r+2k-3i} }{(q^2;q^2)_{r-k} (q^2;q^2)_{k-i}(q^2;q^2)_i }
\end{equation}
Introducing  $r=d_1+d_2+d_3, k=d_2+d_3,  i=d_3$,  we can transform it into
\begin{equation}
    P_{3_1}(x,a,q) = \sum_{d_1,d_2,d_3\geq 0} (-q)^{2d_1d_2+2d_1d_3+2d_2^2+2d_2d_3+3d_3^2} \frac{(x a^2 q^{-2})^{d_1}}{(q^2;q^2)_{d_1}} \frac{(x a^2)^{d_2}}{(q^2;q^2)_{d_2}}\frac{(x a^4 q^{-3})^{d_3}}{(q^2;q^2)_{d_3}}. \label{PxC-31}
\end{equation}
Comparing with the quiver partition function (\ref{PQ}), it is easy to see that $P_{3_1}(x,a,q)=P_Q(\boldsymbol{x},q)$ for
\begin{align}\label{eq:trefoil quiver}
     C &= \left[
\begin{array}{ccc}
0 & 1 & 1 \\
1 & 2 & 2 \\
1 & 2 & 3 \\
\end{array}
\right] \\
\boldsymbol{x}&= \left[x a^2 q^{-2},x a^2 ,x a^4 q^{-3} \right]\label{trefoil-quiver}
\end{align}

\subsection{Physical and geometric interpretations} \label{sec:interpretations}

Let us now take a detour to the physical picture. As was mentioned earlier in the introduction, Ooguri and Vafa \cite{OV9912} considered open topological strings in the resolved conifold with Lagrangian submanifold $L_K$ associated with the knot. After compactification of the resolved conifold, they obtained an effective 3d $\mathcal{N}=2$ theory $T[L_K]$ in which the counts of Bogomol'nyi–Prasad–Sommerfield (BPS) states were new invariants of a knot $K$. Those invariants were further studied by Labastida, Mariño and Vafa in \cite{LM01,LM02,LMV00}, and are commonly called LMOV invariants, abbreviating the last names of the authors.
Mathematically,  LMOV invariants can be defined as the numerical exponents $N_{r,i,j}$ of the product decomposition of the HOMFLY-PT generating function in terms of quantum dilogarithms:
\begin{equation}\label{eq:P_K=00003DExp}
P_{K}(x,a,q)=\prod_{r\geq 1}\prod_{i,j\in \mathbb{Z}}(x^{r} a^i q^j;q^2)_{\infty}^{N_{r,i,j}}\,.
\end{equation}

Taking advantage of the knot-quiver correspondence, we can use the product form (\ref{PQ-product}) in order to relate quiver DT invariants of a symmetric quiver to LMOV invariants $N(x,a,q)=\sum_{r,i,j}N_{r,i,j}x^{r}a^{i}q^{j}$ of a corresponding knot:
\begin{equation}\label{eq:LMOV vs DT}
N(x,a,q)=\left.\Omega(\boldsymbol{x},q)\right|_{x_i=x a^{a_i} q^{q_i-C_{ii}}}.
\end{equation}
For example, in the case of trefoil knot and its quiver (\ref{eq:trefoil quiver})
the linear terms in $x$ are given by
\begin{align}
    \Omega(\boldsymbol{x},q) = & -x_1 - q^2x_2  + q^3x_3 + \dots
\end{align}
Applying the~relation (\ref{eq:LMOV vs DT}), we get the first few non-zero LMOV invariants:
\begin{equation}
    N(x, a, q) =  -a^2 q^{-2} \left(1+a^2 q^5+q^6\right) x + O(x^2).
\end{equation}

Note that in general the computation of DT and LMOV invariants is computationally demanding -- in subsequent sections we will show how to compute them more efficiently, rather than expanding as a series in $x$ and matching the terms order by order. 

DT invariants perspective is crucial for understanding the physical and geometric interpretation of the knot-quiver correspondence. 
Analysing the linear order ($|\boldsymbol{d}|=1$) of the quiver partition function one can see that quiver nodes correspond to the simplest DT invariants; other ones include their composition with a contribution from arrows. This can be rephrased as follows: from the physical perspective quiver nodes correspond to ground states in the BPS spectrum and quiver arrows encode their interactions and lead to the presence of bound states (for more details see \cite{EKL1811}). 

On the other hand, from the M-theory perspective, BPS states corresponding to knots are known to arise from M2-branes that wrap holomorphic disks (for more details see \cite{OV9912}). Similarly to the case of physical interpretation, we can conclude that from the geometric point of view, quiver nodes correspond to the basic holomorphic disks. On the other hand, each pair of arrows geometrically corresponds to a unit of linking between boundaries of holomorphic disks. Broad discussion of this subject can be found in  \cite{EKL1811}.

\section{Equivalent quivers}

\subsection{Unlinking}\label{sec:unlinking}

\begin{figure}[h!]
    \centering
    \includegraphics[width=0.75\textwidth]{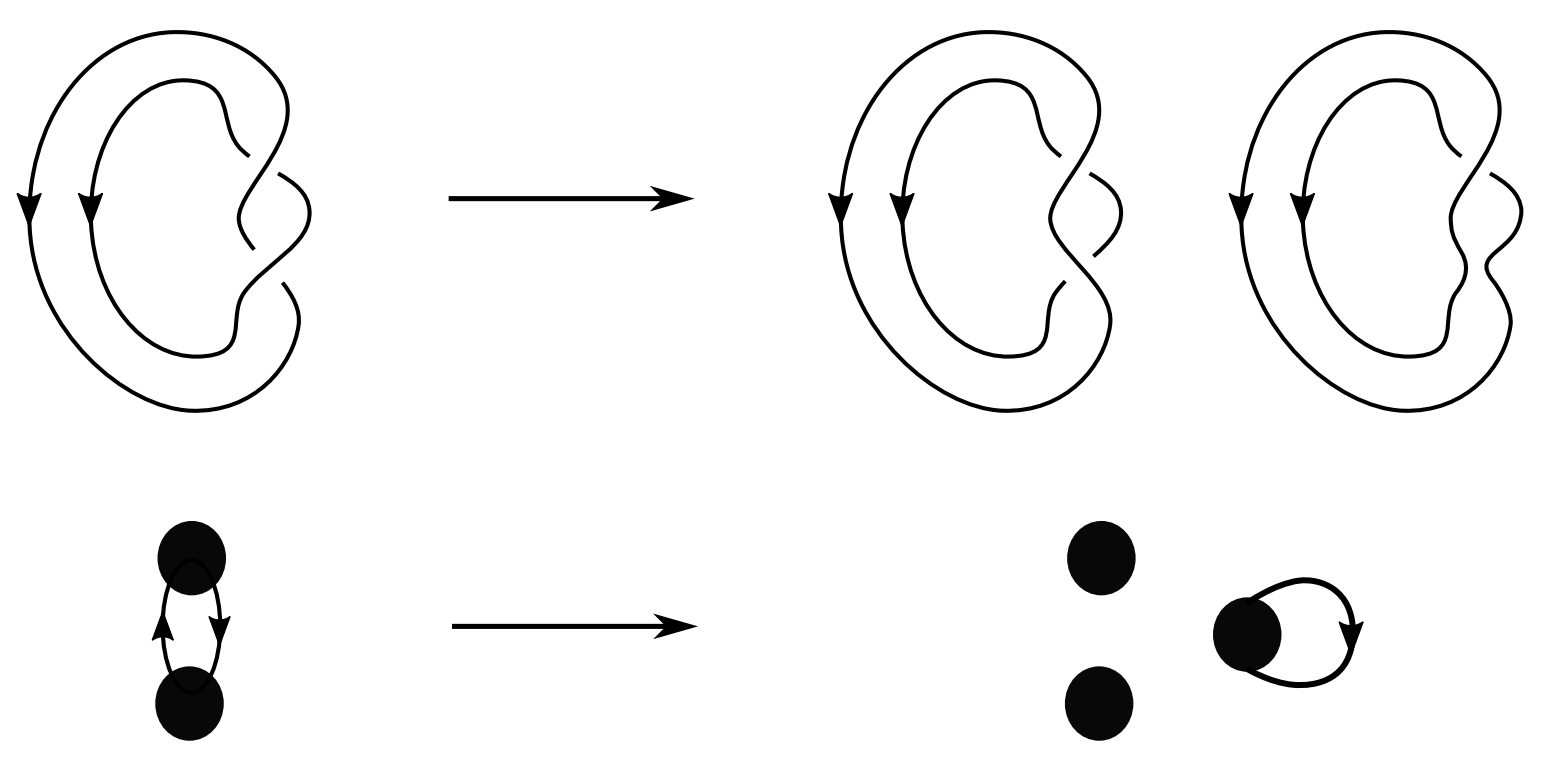}
    \caption{Unlinking in a nutshell: removal of a pair of arrows in a symmetric quiver (as shown on the bottom) amounts to skeining the corresponding pair of holomorphic discs (top), resulting in an extra node and a loop.}
    \label{fig:multi-cover skein relation}
\end{figure}

We can combine the geometric interpretation of the knot-quiver correspondence discussed in the previous section with the idea of skein relations for the boundaries of holomorphic disks \cite{ES1901}. Let's take a look at the example of a pair of disks whose boundaries have linking number $+1$ (Figure \ref{fig:multi-cover skein relation}).

\vspace{1cm}
\vspace{1cm}

We can see that changing and resolving one crossing  transforms initial pair of disks into an unlinked pair and an extra disk with self-linking. Recalling \ref{sec:interpretations}, we can interpret it as a transformation from a quiver with two nodes and one pair of arrows into a quiver with three nodes and one loop. One can check that if we substitute $x_3=q^{-1}x_1 x_2$ (generating parameter of the new node is the product of the initial ones), then the quiver partition function of both quivers is the same \cite{EKL1910}!

We can generalise the transformation described above into an operator that acts on arbitrary quiver adjacency matrix and generating parameters. Namely, for $Q=(C,\boldsymbol{x})$ given by
\begin{equation}
    \begin{split}
    C & =\left[\begin{array}{ccccccc}
 C_{11}  &  \cdots  &  C_{1i}  &  \cdots  &  C_{1j}  &  \cdots  &  C_{1m}\\
   &  \ddots\  &  \vdots &   &  \vdots  &   &  \vdots\\
 &  &  C_{ii}  &  \cdots &  C_{ij}  &  \cdots &  C_{im}\\
 &  &  &  \ddots\  &  \vdots  &   &  \vdots\\
 &  &  &  &  C_{jj}  &  \cdots  &  C_{jm}\\
 &  &  &  &  &  \ddots &  \vdots\\
 &  &  &  &  &  & C_{mm}
\end{array}\right]\,,\\
\boldsymbol{x} & =\left[x_{1},\dots,x_{i},\dots,x_{j},\dots,x_{m}\right]\,,
\end{split}
\end{equation}
we define \textit{unlinking} of distinct nodes $i$ and $j$ as $U(ij)Q = (U(ij)C,U(ij)\boldsymbol{x})$, where
\begin{equation}\label{eq:unlinking definition}
    \begin{split}U(ij)C & =\left[\begin{array}{cccccccc}
 C_{11}  &  \cdots &  C_{1i}  &  \cdots &  C_{1j}  &  \cdots &  C_{1m}  &  C_{1i}+C_{1j}\\
 &  \ddots\  &  \vdots &   &  \vdots &   &  \vdots &  \vdots\\
 &  &  C_{ii}  &  \cdots &  C_{ij}-1  &  \cdots &  C_{im}  &  C_{ii}+C_{ij}-1\\
 &  &  &  \ddots\  &  \vdots &   &  \vdots &  \vdots\\
 &  &  &  &  C_{jj}  &  \cdots &  C_{jm}  &  C_{ij}-1+C_{jj}\\
 &  &  &  &  &  \ddots\  &  \vdots &  \vdots\\
 &  &  &  &  &  &  C_{mm}  &  C_{im}+C_{jm}\\
 &  &  &  &  &  &    &  C_{ii}+C_{jj}+2C_{ij}-1 
\end{array}\right]\,,\\
U(ij)\boldsymbol{x} & =\left[x_{1},\dots,x_{i},\dots,x_{j},\dots,x_{m},q^{-1}x_{i}x_{j}\right]\,.
\end{split}
\end{equation}
The crucial property of unlinking is the preservation of the quiver partition function \cite{EKL1910}:
\begin{equation}
    P_Q(\boldsymbol{x},q) = P_{U(ij)Q}(U(ij)\boldsymbol{x},q)\,.
\end{equation}
There exist also other transformations of quivers -- linking and addition of a redundant pair -- that keep the quiver partition function invariant; information about them can be found in \cite{EKL1910}.

\subsection{Permutohedra graphs}\label{sec:Permutohedra graphs}

In previous section we saw that unlinking preserves the quiver partition function, but it produces an extra node. However, if two different quivers can be unlinked to the same one, both their size and quiver partition function are equal -- we call those quivers \textit{equivalent}. Let us look at the simple example of two equivalent quivers corresponding to the figure-eight knot:

\begin{equation}\label{eq:figure-eight quivers}
\begin{split}
    C_1 & =\left[\begin{array}{ccccc}
    0  &- 1 & -1 &  0 &  0\\
    -1 & -2 & -2 & -1 &  \mathbf{-1}\\
    -1 & -2 & -1 &   \mathbf{0} &  0\\
    0  & -1 &   0 &  1 &  1\\
    0  &  -1 &  0 &  1 &  2
    \end{array}\right]\,, \qquad  \qquad 
     C_2=\left[\begin{array}{ccccc}
    0  &- 1 & -1 &  0 &  0\\
    -1 & -2 & -2 & -1 &  \mathbf{0}\\
    -1 & -2 & -1 &  \mathbf{-1} & 0\\
    0  & -1 &  -1 &  1 & 1\\
    0  &   0 &  0 &  1 & 2
    \end{array}\right] \,,\\
    \boldsymbol{x_1} & = \left[ 1, a^{-2} q^2 , q^{-1} , q, a^2 q^{-2} \right]
     = \boldsymbol{x_2}  \,.
\end{split}
\end{equation}
If we apply unlinking  $U(34)$  to $Q_1$ and $U(25)$ to $Q_2$, in both cases we obtain
\begin{equation}
\begin{split}
   U(34) C_1 = U(25) C_2 &= \left[\begin{array}{cccccc}
        0  &- 1 & -1 &  0 &  0 & -1 \\
        -1 & -2 & -2 & -1 &  -1 & -3 \\
        -1 & -2 & -1 &  -1 &  0 & -2 \\
        0  & -1 &  -1 &  1 &  1 & 0 \\
        0  &  -1 &  0 &  1 &  2 & 1 \\
        -1 & -3 & -2 & 0 & 1 & -1 
        \end{array}\right] \ , \\
    U(34) \boldsymbol{x_1} = U(25) \boldsymbol{x_2} & = \left[ 1, a^{-2} q^2 , q^{-1} , q, a^2 q^{-2} , 1 \right] \,.
\end{split}
\end{equation}
We can represent this structure in a form of a graph whose nodes are quivers $Q_1$ and $Q_2$, and the edge represents the equivalence relation defined by the possibility of unlinking to the same quiver \cite{JKLNS2105}:

\[\begin{tikzcd}
	{Q_1} && {Q_2}
	\arrow[no head, from=1-1, to=1-3]
\end{tikzcd}\]

\begin{figure}[h!]
    \centering
 \input{Figures/permut.tikz}
    \caption{Examples of permutohedra $\Pi_n$, shown as planar graphs.}
    \label{fig:permutohedra}
\end{figure}
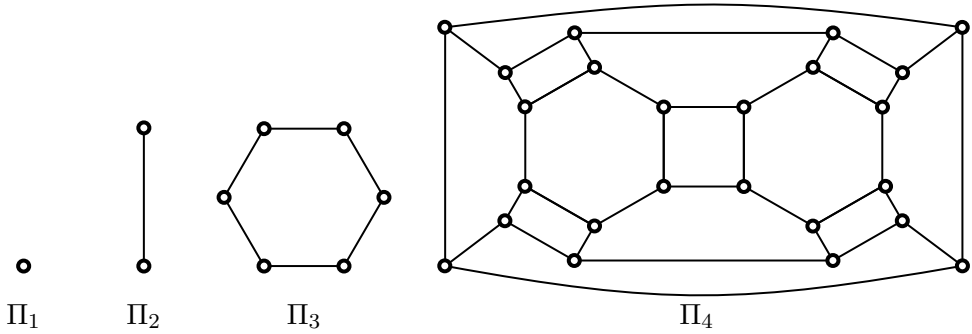

This graph is the simplest nontrivial permutohedron -- a graphical representation of the permutation group in which each node corresponds to a permutation and each edge to the transposition of neighbouring elements. Permutohedra for permutations of 1,2,3 and 4 elements are shown in Figure \ref{fig:permutohedra}.

\begin{figure}[h!]
\centering
\includegraphics[scale=1.5]{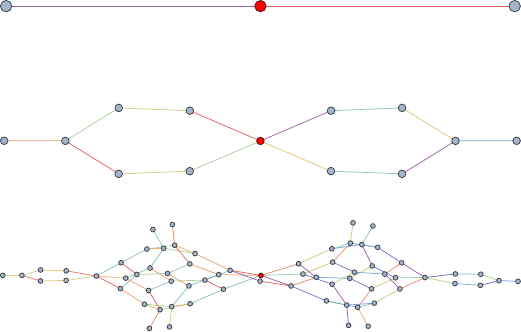}
\caption{Permutohedra graphs for knots $5_1,7_1,9_1$ (from top to bottom). Every vertex corresponds to an equivalent quiver, while the edge between two vertices corresponds to a transposition of a pair of arrows (different colours correspond to different transpositions).}
\label{fig:permutohedra examples}
\end{figure}

For the trefoil knot, analogous graph is trivial and consists of only one vertex.
Graphs representing the structure of equivalent quivers for torus knots $5_1$, $7_1$ and $9_1$ are presented in Figure \ref{fig:permutohedra examples}, and we can see that they are made of several permutohedra glued together \cite{JKLNS2105}. This is a general feature and because of that those structures are called \textit{permutohedra graphs}; for more details see  \cite{JKLNS2105}.

\subsection{Permutohedra from unlinking}

We can understand the role of permutations for equivalent quivers if we notice that  $Q_1$ and $Q_2$ from \eqref{eq:figure-eight quivers}  can be obtained in the process based on unlinking nodes of the following quiver:
\begin{equation}\label{eq:Enlarged prequiver 4_1}
\begin{split}
    \check{C}&=\left[\begin{array}{ccc:c}
    0 & -1 & 0 & 0\\
    -1 & -2 & -1 & 1\\
    0 & -1 & 1 & 1\\
    \hdashline
    0 & 1 & 1 & 0
    \end{array}\right]\ ,\\
    \boldsymbol{\check{x}} &  = 
   [ x , x a^{-2} q^2 ,x q ;   a^2q^{-2}] \,.
\end{split}
\end{equation}
 \textit{in different order} \cite{KLNS2312}. Namely, if we apply unlinking $U(24)$ and then $U(34)$, we obtain
\begin{equation}
U(34)U(24)\check{C}=
\left[\begin{array}{ccccc:c}
    0  &- 1 & -1 &  0 &  0 &  0\\
    -1 & -2 & -2 & -1 &  -1 &  0\\
    -1 & -2 & -1 &   0 &  0 &  0\\
    0  & -1 &   0 &  1 &  1 &  0\\
    0  &  -1 &  0 &  1 &  2 &  0\\
    \hdashline
    0  & 0 & 0 &  0 &  0 &  0
    \end{array}\right]\,.
\end{equation}
The other permutation gives
\begin{equation}
U(24)U(34)\check{C}=
\left[\begin{array}{ccccc:c}
    0  &- 1 & -1 &  0 &  0 &  0\\
    -1 & -2 & -2 & -1 &  0 &  0\\
    -1 & -2 & -1 &  -1 &  0 &  0\\
    0  & -1 &  -1 &  1 &  1 &  0\\
    0  &  0 &  0 &  1 &  2 &  0\\
    \hdashline
    0  & 0 & 0 &  0 &  0 &  0
    \end{array}\right]\,,
\end{equation}
and in both cases the generating parameters are given by
\begin{equation}
\left[ x, x a^{-2} q^2 ,x q^{-1} , xq, xa^2 q^{-2} ;   a^2q^{-2} \right] \,.
\end{equation}
Note that the last row and column contain only zeroes, so the corresponding BPS state does not interact with the rest and in both cases contributes the overall factor $(a^2q^{-2};q^2)^{-1}_{\infty}$  to the quiver partition function. Division by this factor corresponds to erasing the last node (and corresponding row and column), leading to  $Q_1$ and $Q_2$ from \eqref{eq:figure-eight quivers} .  Summing up, we can see that permutohedron structure comes from performing unlinking in different order. This is a general feature, for more details see \cite{KLNS2312}.

\section{Quiver diagonalization}

\subsection{\texorpdfstring{$m$-loop quivers}{m-loop quivers}}\label{sec:m-loop_quivers}

Among all symmetric quivers, a special role is played by  \emph{$m$-loop quivers}, which consist of only a single node and $m$ loops around this node

\begin{equation}
\begin{tabular}{ccc}
    \vspace{1cm} \\
    $Q = \qquad\quad$
    &
    \begin{tikzpicture}[scale=2, every loop/.style={}, baseline=(current  bounding                           box.center),overlay]
        \draw [fill] (0, 0) circle [radius=0.02];
        \node [draw=none] {} edge [in=50-20,out=130+20,loop,scale=5] ();
        \node [draw=none] {} edge [in=50-10,out=130+10,loop,scale=3.5] ();
        \node [draw=none] {} edge [in=50+20,out=130-20,loop,scale=2] ();
        \node at (0.27, 0.5) { $\cdots$ };
    \end{tikzpicture}
    & $\qquad, \qquad\quad C=[\,m \,]\,$.
\end{tabular}
\end{equation}
Their quiver generating series is given by
\begin{equation}\label{eq:m-loop}
P_{\text{$m$-loop}}(x,q) = 
\sum_{d=0}^{\infty} \frac{(-q)^{md^2}}{(q^2;q^2)_d}\,x^d
=
1 + (-1)^m\frac{q^m}{1-q^2}x + \frac{q^{4m}}{(1-q^2)(1-q^4)}x^2 + O(x^3).
\end{equation}
For $m=0$ and $m=1$ one can recognise the well-known $q$-identities: \footnote{For a comprehensive list of $q$-identities see, for example, \hyperlink{https://dlmf.nist.gov/17.2}{NIST Digital Library of Mathematical Functions}.}
\begin{equation}\label{eq:one and two loop quivers}
    \sum_{d=0}^{\infty} \frac{1}{(q^2;q^2)_d}\,x^d = \frac{1}{(x;q^2)_{\infty}}, \qquad \sum_{d=0}^{\infty} \frac{(-q)^{d^2}}{(q^2;q^2)_d}\,x^d = 
    (qx;q^2)_{\infty}.
\end{equation}
Using the plethystic exponential operator and the language of DT invariants, one gets
\begin{equation}
    \Omega_{\text{0-loop}}(x,q) = -x,\quad \Omega_{\text{1-loop}}(x,q) = qx\, .
\end{equation}
It it tempting to generalise these results for $m>1$, but then the quiver partition function does not admit a finite factorisation into a number of infinite $q$-Pochhammers à la (\ref{eq:one and two loop quivers}). In general, for an $m$-loop quiver with $m>1$ the structure of the product form, or, equivalently, the DT generating function $\Omega(x,q)$, is much more involved. One can use different methods to compute DT invariants of an $m$-loop quiver -- e.g. recursion relations. They have not just one, but several very interesting (but quite technical) combinatorial interpretations -- we will not give details here, referring the reader to the original sources \cite{Rei12,KS1608,DotsenkoFeiginReineke}.
Some examples can be computed directly (e.g. in Mathematica) up to a fixed order in $x$, and are shown in Table~\ref{tab:mqPoch}.
\begin{table}[h!]
    \centering
    \renewcommand{\arraystretch}{1.5}
    \begin{tabular}{||c|c||}
    \hline
$m$ &\ $\Omega_{\text{$m$-loop}}(x,q)$ \\
\hline
\hline
0 &\ $-x$ \\
\hline
1 &\ $qx$ \\
\hline
2 &\ $-q^2x + q^4x^2 - q^8x^3 + O(x^4)$ \\
\hline
3 &\ $q^3x + q^8x^2 + (q^{11}+q^{13}+q^{17})\, x^3 + O(x^4)$ \\
\hline
4 &\ $-q^4x + (q^8+q^{12})x^2 - (q^{14}+q^{16}+q^{18}+q^{20}+q^{22}+q^{26})\, x^3
+ O(x^4) $ \\
\hline
\end{tabular}
    \caption{DT invariants of $m$-loop quivers.}
    \label{tab:mqPoch}
\end{table}

From the perspective of knots and their invariants, $m$-loop quiver partition function encodes the reduced coloured HOMFLY-PT invariants of a framed unknot, with framing being equal to the number of loops $m$ \cite{KRSS1707long}. 
Another equivalent interpretation is that of open topological string amplitudes for branes in $\mathbb{C}^3$ geometry \cite{MarCS}.

What will be most relevant in our context, $m$-loop quivers serve as building blocks of more complicated symmetric quivers. In order to show that, one has to use the unlinking operation. In the following Section we discuss how to bring any quiver to an equivalent diagonal form, so that their DT invariants are expressed entirely in terms invariants of $m$-loop quivers.

\subsection{Approximations and infinite limit}

In Section \ref{sec:m-loop_quivers}  we saw quivers that contain only loops as the simplest case where DT invariants can be computed order by order in $x$. One may then ask what happens if we start with an arbitrary symmetric quiver and use unlinking to delete every arrow connecting different nodes. In fact, we have already seen it in Section \ref{sec:unlinking}, when the quiver with two nodes and one pair of arrows was transformed into a quiver with three unconnected nodes (one of which has a loop). Recalling  Table \ref{tab:mqPoch}, we can immediately see that in this case we have 3 non-zero DT invariants. However, this example is exceptionally simple, so we move our attention to its slight generalisation:
\begin{equation}\label{eq:A3}
Q=\begin{tikzcd}
	{\bullet} & {\bullet} & {\bullet}
	\arrow[curve={height=-6pt}, from=1-2, to=1-3]
	\arrow[curve={height=-6pt}, from=1-3, to=1-2]
	\arrow[curve={height=-6pt}, from=1-1, to=1-2]
	\arrow[curve={height=6pt}, tail reversed, no head, from=1-1, to=1-2]~
\end{tikzcd},
\qquad
C=
\begin{bmatrix}
0 & 1 & 0 \\
1 & 0 & 1 \\
0 & 1 & 0 \\
\end{bmatrix},
\qquad
\boldsymbol{x} = \left[
\begin{array}{c}
x_1\\
x_2 \\
x_3 \\
\end{array} \right].
\end{equation}
Let us get rid of those two pairs of arrows by applying unlinking $U(12)$ and $U(23)$:
\begin{equation}\label{eq:A3_sequencea}
\begin{tikzpicture}[
    node distance=1mm and 0mm,
    baseline]
\matrix (M1) [matrix of nodes,{left delimiter=[},{right delimiter=]}]
{
 0 & \textcolor{red}{1} & 0 \\
 \textcolor{red}{1} & 0 & 1 \\
 0 & 1 & 0 \\
};
\draw[black,dashed] 
        (M1-1-1.north west) -| (M1-3-3.south east) -| (M1-1-1.north west);
\end{tikzpicture}
\ \rightarrow \
\begin{tikzpicture}[
    node distance=1mm and 0mm,
    baseline]
\matrix (M1) [matrix of nodes,{left delimiter=[},{right delimiter=]}]
{
 0 & 0 & 0 & 0 \\
 0 & 0 & \textcolor{red}{1} & 0 \\
 0 & \textcolor{red}{1} & 0 & 1 \\
 0 & 0 & 1 & 1 \\
};
\draw[black,dashed] 
        (M1-1-1.north west) -| (M1-3-3.south east) -| (M1-1-1.north west);
\end{tikzpicture}
\ \rightarrow \
\begin{tikzpicture}[
    node distance=1mm and 0mm,
    baseline]
\matrix (M1) [matrix of nodes,{left delimiter=[},{right delimiter=]}]
{
 0 & 0 & 0 & 0 & 0 \\
 0 & 0 & 0 & 0 & 0 \\
 0 & 0 & 0 & 1 & 0 \\
 0 & 0 & 1 & 1 & 1 \\
 0 & 0 & 0 & 1 & 1 \\
};
\draw[black,dashed] 
        (M1-1-1.north west) -| (M1-3-3.south east) -| (M1-1-1.north west);
\end{tikzpicture}
\end{equation}
Now the subquiver corresponding to dashed region of the matrix is diagonal and one can check that its quiver partition function agrees with \eqref{eq:A3}  up to $O(\boldsymbol{x}^2)$ \cite{JKLNS2212}.\footnote{When we consider the order in $x$, we discard all subscirpts and trea all $x_1,x_2,\dots$  as $x$.} However, we created new nodes that are connected by new arrows -- we can get rid of them by applying unlinking $U(34)$ and $U(45)$:
\begin{equation}
\begin{aligned}
\begin{tikzpicture}[
    node distance=1mm and 0mm,
    baseline]
\matrix (M1) [matrix of nodes,{left delimiter=[},{right delimiter=]}]
{
 0 & 0 & 0 & 0 & 0 \\
 0 & 0 & 0 & 0 & 0 \\
 0 & 0 & 0 & \textcolor{red}{1} & 0 \\
 0 & 0 & \textcolor{red}{1} & 1 & 1 \\
 0 & 0 & 0 & 1 & 1 \\
};
\draw[black,dashed] 
        (M1-1-1.north west) -| (M1-5-5.south east) -| (M1-1-1.north west);
\end{tikzpicture}
\ \rightarrow \
\begin{tikzpicture}[
        node distance=1mm and 0mm,
    baseline]
\matrix (M1) [matrix of nodes,{left delimiter=[},{right delimiter=]}]
{
 0 & 0 & 0 & 0 & 0 & 0 \\
 0 & 0 & 0 & 0 & 0 & 0 \\
 0 & 0 & 0 & 0 & 0 & 0 \\
 0 & 0 & 0 & 1 & \textcolor{red}{1} & 1 \\
 0 & 0 & 0 & \textcolor{red}{1} & 1 & 1 \\
 0 & 0 & 0 & 1 & 1 & 2 \\
};
\draw[black,dashed] 
        (M1-1-1.north west) -| (M1-5-5.south east) -| (M1-1-1.north west);
\end{tikzpicture}
\ \rightarrow \
\begin{tikzpicture}[
        node distance=1mm and 0mm,
    baseline]
\matrix (M1) [matrix of nodes,{left delimiter=[},{right delimiter=]}]
{
 0 & 0 & 0 & 0 & 0 & 0 & 0 \\
 0 & 0 & 0 & 0 & 0 & 0 & 0 \\
 0 & 0 & 0 & 0 & 0 & 0 & 0 \\
 0 & 0 & 0 & 1 & 0 & 1 & 1 \\
 0 & 0 & 0 & 0 & 1 & 1 & 1 \\
 0 & 0 & 0 & 1 & 1 & 2 & 2 \\
 0 & 0 & 0 & 1 & 1 & 2 & 3 \\
};
\draw[black,dashed] 
        (M1-1-1.north west) -| (M1-5-5.south east) -| (M1-1-1.north west);
\end{tikzpicture}
\end{aligned}
\end{equation}
Now the top-left $5\times 5$ dashed subquiver is diagonal and its quiver partition function agrees with \eqref{eq:A3}  up to $O(x^3)$ \cite{JKLNS2212}. However, we created new nodes and new arrows again, so we can apply appropriate unlinkings to get rid of them. This process gradually gives better and better approximations of the initial quiver partition function. In the infinite limit we get a (usually infinite) quiver that contains only nodes and loops, whose partition function is equal to the initial one \cite{JKLNS2212}. Since the adjacency matrix of the final quiver is diagonal, the whole procedure is called \textit{diagonalization} -- more details can be found in \cite{JKLNS2212}.

\subsection{Computation of DT invariants}

Since we know the DT invariants of $m$-loop quivers (see Table \ref{tab:mqPoch}), the first $n$ steps of diagonalization procedure can be used to compute DT invariants up to $O(\boldsymbol{x}^{n+1})$ which is the most efficient method developed so far.

We can see how it works on the trefoil example.
In this case the~adjacency matrix and the~change of variables are given by \cite{KRSS1707long}
\begin{equation}\label{eq:trefoil_quiver_matrix}
    C=\left[
\begin{array}{ccc}
 0 & 1 & 1 \\
 1 & 2 & 2 \\
 1 & 2 & 3 \\
\end{array}
\right]~,
\qquad
\left[
\begin{array}{c}
x_1 \\
x_2 \\
x_3 \\
\end{array} \right]  = \left[
\begin{array}{c}
x a^2 q^{-2}\\
x a^2 \\
x a^4 q^{-3} \\
\end{array} \right].
\end{equation}
The application of unlinkings $U(12),U(13),U(23)$ leads to
\begin{equation} 
\begin{tikzpicture}[
    node distance=1mm and 0mm,
    baseline]
\matrix (M1) [matrix of nodes,{left delimiter=[},{right delimiter=]}]
{
0 & 0 & 0 & 0 & 0 & 0 & 0 \\
 0 & 2 & 0 & 2 & 2 & 3 & 2 \\
 0 & 0 & 3 & 3 & 3 & 4 & 3 \\
 0 & 2 & 3 & 3 & 3 & 5 & 5 \\
 0 & 2 & 3 & 3 & 4 & 5 & 5 \\
 0 & 3 & 4 & 5 & 5 & 8 & 7 \\
 0 & 2 & 3 & 5 & 5 & 7 & 6 \\
};
\draw[black,dashed] 
        (M1-1-1.north west) -| (M1-3-3.south east) -| (M1-1-1.north west);
\end{tikzpicture}~.
\end{equation}
Since the~seven diagonal entries of this matrix are the~only ones which contribute to quadratic terms $x_ix_j$ in $P_Q(\boldsymbol{x},q)$, we do not have to perform the next step in full form, but we can conclude that up to $O(\boldsymbol{x}^{3})$, the diagonal matrix approximating $C$ equals  $\text{diag}\big(0, 2, 3 ; 3, 4, 8, 6\big)$, whereas the change of variables is given by $\left(x_1,x_2,x_3 ; q^{-1}x_1x_2,q^{-1}x_1x_3,q^{-1}x_2x_3,q^{-1}x_2x_3\right)$.
Following  Section \ref{sec:interpretations} and expanding the~product of $C_{ii}$-loop quivers, we obtain DT invariants:
\begin{align}
    \Omega(\boldsymbol{x},q) = & -x_1 - q^2x_2  + q^3x_3 \\
    & + q^2x_1x_2  - q^3x_1x_3 + q^4x_2^2 - q^5x_2x_3  - q^7x_2x_3 + q^8x_3^2 + O\left(\boldsymbol{x}^3\right)~. \nonumber
\end{align}
The application of the~change of variables \eqref{eq:trefoil_quiver_matrix} translates them into LMOV invariants:
\[
    N(x, a, q) =  -a^2 q^{-2} \left(1+a^2 q^5+q^6\right) x + a^4 q \left(a^2+q\right) \left(1+q^6+a^2 q^7\right) + O\left(x^{3}\right)~.
\]

Summing up, the above method turns out to be very efficient in determining the combinatorial structure among the DT invariants of symmetric quivers, and can be translated to the LMOV invariants via the knot-quiver correspondence. As such, LMOV invariants also turn into purely combinatorial objects which can be counted almost directly from the quiver adjacency matrix via unlinking operation.
We invite the reader to learn more related examples in \cite{JKLNS2212}.

\section{\texorpdfstring{$F_K$}{FK} invariants and quivers for knot complements}

\subsection{Overview}

Knot theory, apart from being an extensive branch of mathematics on its own, is also an integral part of the low-dimensional topology in general (i.e. a study of manifolds in dimensions three and four). The classical result by Lickorish and Wallace from 1960's states that every closed compact 3-manifold can be obtained as a surgery \footnote{Dehn surgery is an operation on manifolds which consists of two steps: 1) cutting out a tubular neighbourhood of a link, and 2) gluing it back, but with a more general homeomorphism identifying the boundary tori with the gluing piece.} on some link in $S^3$. Among them, a special class of manifolds is obtained from a surgery on some knot -- for example, the family of lens spaces is obtained from surgeries on an unknot in $S^3$.
As a result, invariants of such closed 3-manifolds are closely related to invariants of knot complements.
In this section we focus on one such invariant, which recently has gained quite a lot of attention. As we shall see below, it also enjoys a form of the knot-quiver correspondence.

\begin{figure}[ht]
    \centering
    \includegraphics[width=0.5\linewidth]{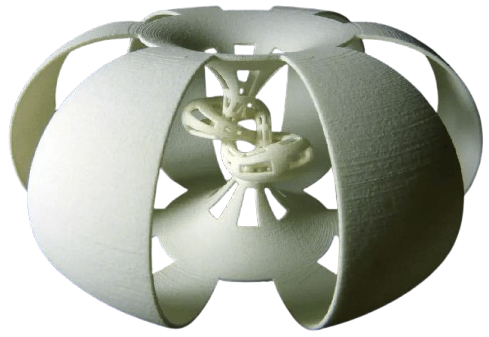}
    \caption{A real-life model of $S^3\setminus 4_1$ by Henry Segerman.}
    \label{fig:figure-eight_complement}
\end{figure}

To begin with, let $K$ be a knot in $S^3$; define knot complement as $S^3\setminus [K]$, where $[K]$ is an open tubular neighbourhood of $K$. For simplicity, it is often written as $S^3\setminus K$.
Despite the fact that boundary of $S^3\setminus K$ is homeomorphic to a torus, the topology of the complement itself is dictated by the topology of a knot and can be quite non-trivial. If $K$ is an unknot, then both $S^3\setminus \bigcirc$ and $[\bigcirc]$ are simply a solid torus.
Another example is the complement of figure-eight knot, shown in Figure \ref{fig:figure-eight_complement}.\footnote{\textcolor{blue}{https://www.thingiverse.com/henryseg/designs}}

Our goal is to define a two-variable series for knot complements which we denote $F_K(x,q)$, starting from a sequence of coloured Jones polynomials $J^K_n(q)$ in the reduced normalisation\footnote{To be more precise, here we consider the $SU(2)$ version of $F_K$. It is also possible to incorporate HOMFLY-PT variable $a$ into $F_K$, i.e. construct an invariant $F_K(x,a,q)$, but we will not go into the details here.}. Let $q=e^{\hbar}$. The starting point is Melvin-Morton-Rozansky (MMR) asymptotic expansion \cite{rozansky1997higher} which is a double series expansion of $J_n(e^{\hbar})$:
\begin{equation}
J_n(e^{\hbar}) = \frac{1}{\Delta_K(x)} + \frac{P_1(x)}{\Delta_K(x)^3}\hbar + \frac{P_2(x)}{\Delta_K(x)^5}\hbar^2 + \dots  = \sum_{m=0}^\infty \sum_{j=0}^m c_{m,j}n^j\hbar^m.
\end{equation}
(Here $x:=q^n$.)
This expansion is very remarkable on its own, as it relates three very different knot invariants -- the coloured Jones polynomial, Alexander polynomial $\Delta(x)$, and finite-type Vassiliev invariants $c_{m,j}$ (more about them in \cite{jackson2019introduction}). 
The main trick is to apply the \emph{Borel resummation} to the above series, which produces the right hand side as a function of $q$ and $x:=q^n$:
\begin{equation}\label{eq:Borel resum}
\sum_{m=0}^\infty \sum_{j=0}^m c_{m,j}n^j\hbar^m \quad 
\stackrel{\text{\emph{Borel resum}}}{=} F_K(x,q)(x^{1/2}-x^{-1/2})^{-1},
\end{equation}
This definition should, at least in principle, work for every knot. Unfortunately, the Borel resummation technique can be very tricky, and in practice it is often better to use other methods, such as recursion relations or surgery formulae. While the latter requires more technical explanations, the recursion relations can be stated in a simpler way and can be easily solved, e.g., in Mathematica:
\begin{coj}\cite{GM1904,EKL2108}
    For any knot $K\subset S^3$, the series $f_K(x,q):=F_K(x,q)(x^{1/2}-x^{-1/2})^{-1}$
    are solutions to the $q$-difference equation
    \begin{equation}\label{eq:quantum A-poly}
        \widehat{A}(\widehat{x},\widehat{y})f_K(x,q) = 0
    \end{equation}
\end{coj}
with $\widehat{x}\equiv x$ and $\widehat{y}\widehat{x} = q\widehat{x}\widehat{y}$.
The operator $\widehat{A}(\widehat{x},\widehat{y})$ is called the \emph{quantum A-polynomial of a knot $K$}. Some examples on how to compute such operators are described in \cite{GS12}.
Therefore, knowing the recursion relation plus a suitable initial condition \cite{Ekholm:2020lqy}, one can compute the series $f_K(x,q)$ (likewise, $F_K(x,q)$) up to any given order in $x$.

Let's now discuss some examples. The simplest case is a 0-framed unknot, whose complement is homeomorphic to a solid torus. We know that all coloured Jones polynomials are equal to 1. Since there is no dependence on $q$, there is no need to perform the Borel resummation in (\ref{eq:Borel resum}), and we get the answer immediately:
\begin{equation}\label{eq:F_K unknot}
    F_{\bigcirc}(x,q) = x^{1/2}-x^{-1/2};\quad F^+_{\bigcirc}(x,q) = x^{1/2}.
\end{equation}
On the other hand, in more complicated cases it is useful to invoke the relation (\ref{eq:quantum A-poly}). For example, the quantum $\widehat{A}$-polynomial for figure-eight knot $4_1$ is given by \cite{GS12}:
\begin{equation}
\widehat{A}(\widehat{x},\widehat{y}) = q^3(1 - q^6 \widehat{x}^4) \widehat{x}^4 - (1 - q^4 \widehat{x}^4) (1 - q^2 \widehat{x}^2 - (q^2 + q^6) \widehat{x}^4 - q^6 \widehat{x}^6 + q^8 \widehat{x}^8) \widehat{y} + q^5 (1 - q^2 \widehat{x}^4) \widehat{x}^4 \widehat{y}^2.
\end{equation}
Solving (\ref{eq:quantum A-poly}) order by order in $x$ gives
\begin{equation}\label{eq:F_K plus}
    F^+_{4_1}(x,q) = x^{1/2} + 2x^{3/2} + (q^{-1}+3+q)x^{5/2} + (2q^{-2}+2q^{-1}+5+2q+2q^2)x^{7/2} + \dots.
\end{equation}
As conjectured in \cite{GM1904}, $F_K(x,q)$ is a topological invariant of $S^3\setminus K$, where $x$ is identified with the meridian generator along its toric boundary (it is the same $x$ which enters the knot A-polynomial $A(x,y)$ \cite{GukovSaberi}). Note that in the case of HOMFLY-PT generating series $x$ played a completely different role -- the distinction between the two cases is summarised in \cite{Kuch2005}.
To date, the conjecture is confirmed for a variety of knots and links \cite{Park2004,Ekholm:2020lqy,Ekholm:2021irc}. Our next step is to show that $F_K(x,q)$ satisfies a version of the knot-quiver correspondence \cite{Kuch2005,Ekholm:2021irc}.

\subsection{\texorpdfstring{$F_K$}{FK} invariants and quivers}

In order to formulate the knot-quiver correspondence for knot complements,
one can use a direct approach -- matching quiver adjacency matrix and
the change of variables against order by order expansion in $x$. 
However, one important feature of $F_K(x,q)$ is that it splits into two parts related by Weyl symmetry $x \leftrightarrow x^{-1}$:
\begin{equation}
    F_{K}(x,q) = \frac{1}{2}(F^-_K(x,q)+F^+_K(x,q)),
\end{equation}
where $F^-_K(x,q):=-F^+_K(x^{-1},q)$. Since it is a double-ended series in $x$ as well as $x^{-1}$, in order to match it with the quiver generating series, we need to choose either half.
In what follows, we will identify quiver form for the \emph{normalised positive part} of $F_K$:
\begin{equation}\label{eq:knot complement KQ}
    x^\Delta\frac{F^+_{K}(x,q)}{x^{1/2}-x^{-1/2}} \quad \equiv \quad P_Q(x_1,\dots,x_m;q^{1/2})
\end{equation}
for some choice of normalisation exponent $\Delta$\footnote{$\Delta$ is chosen so that the right hand side of (\ref{eq:knot complement KQ}) has the lowest $x$-degree 1.},
where the quiver variables $x_i$ specialise to monomials in $x,q$ (with one crucial distinction that $x_i$ now allowed to have higher degrees of $x$). The~ choice of $q^{1/2}$ instead of $q$ is a matter of convention -- we keep it consistent with \cite{GM1904}.
Therefore, we take (\ref{eq:knot complement KQ}) as a working definition for quivers associated to knot complements.
The choice of $F^+_{K}(x,q)$ is due to the fact that quiver partition function are normally a series in positive powers of $x$, which makes it not possible to combine both positive and negative parts of $F_K$ in the same quiver generating function. Moreover, dividing by the factor $x^{1/2}-x^{-1/2}$ turns on the reduced normalisation for $F_K$. As a result, the quiver in question is associated to the function directly involved in the Borel summation (\ref{eq:Borel resum}), which makes it more convenient for making a contact with the usual HOMFLY-PT quivers.
In particular, \cite{Ekholm:2021irc} suggests that the size of a knot complement quiver is given by $|Q_K|+1$, where $Q_K$ is the usual HOMFLY-PT quiver for the same knot.

We can start with the simplest example, that is a $0$-framed unknot. We have $|Q_K|=1$, so that for $F_K$ we expect a quiver with two nodes.
Consider quiver
\begin{equation}\label{eq:unknot complement quiver}
C=
\begin{bmatrix}
0 & 0 \\
0 & 1 \\
\end{bmatrix}    ,
\quad
\boldsymbol{x} = 
\begin{bmatrix}
    x \\
    q^{1/2}x \\
\end{bmatrix}.
\end{equation}
We can write its partition function immediately in the product form, taking the advantage of the results from Section \ref{sec:m-loop_quivers}:
\begin{equation}
    P_{C}(x_1,x_2;q^{1/2}) = \frac{(qx;q)_{\infty}}{(x;q)_{\infty}} = \frac{1}{1-x}.
\end{equation}
Quite remarkably, under this specialisation of $x_1,x_2$, the dependence on $q$ goes away completely.
We thus get (taking into account (\ref{eq:F_K unknot}))
\begin{equation}
    \left.P_{C}(x_1,x_2;q^{1/2})\right.\vert_{x_1=x,x_2=q^{1/2}x} \, = \, x^{-1}\frac{F^+_{\bigcirc}(x,q)}{x^{1/2}-x^{-1/2}}.
\end{equation}
Therefore, we identified the quiver for the unknot complement $S^3\setminus \bigcirc$ to be
\begin{equation}
\begin{tikzcd}
	{\bullet_{x}} & {\bullet_{qx}}
	\arrow[from=1-2, to=1-2, loop, in=55, out=125, distance=10mm]
\end{tikzcd}
\end{equation}
%


Let us now move on to the case of 0-framed figure-eight knot. $|Q_K|=5$, so that for $F_K$ we expect a quiver with six nodes.
All we need to do is to use finitely many terms in (\ref{eq:F_K plus}) and compare the two sums:
\begin{equation}\label{eq:F_K plus, normalised}
    \frac{F^+_{4_1}(x,q)}{x^{1/2}-x^{-1/2}} = x(1 + 3 x + (q^{-1}+6+q)x^2 + (2q^{-2}+3q^{-1}+11+2q^2+3q)x^3 + O(x^4)),
\end{equation}
\begin{equation}
    P_{Q}(x_1,\dots,x_6;q^{1/2}) = 1 + \frac{\sum_{i=1}^6 (-q^{1/2})^{C_{ii}}x_i}{1-q} + 
    \frac{\sum_{i=1}^6 (-q^{1/2})^{4C_{ii}}x_i^2}{(1-q)(1-q^2)} + 
    \frac{\sum_{\substack{i\neq j}} (-q^{1/2})^{C_{ij}}x_ix_j}{(1-q)^2} + \dots
\end{equation}
Note that quiver partition function always starts from 1, however, this is not true for the $F_K$ invariant. We thus need to multiply (\ref{eq:F_K plus, normalised}) by $x^{\Delta}=x^{-1}$. Then, the first two terms match, and the next terms to compare are:
\begin{equation}
    3 x = \frac{(-q^{1/2})^{C_{11}}x_1+\dots+(-q^{1/2})^{C_{66}}x_6}{1-q}.
\end{equation}
Here we can first assume that $x_1=x_2=x_3=x$ with $C_{11}=C_{22}=C_{33}=0$. The remaining condition is $(-q)^{C_{44}}x_4+(-q)^{C_{55}}x_5+(-q)^{C_{66}}x_6=-3qx$, which is satisfied for $C_{44}=C_{55}=C_{66}=1$, $x_4=x_5=x_6=q^{1/2}x$. Doing a similar analysis for the terms with $x^2$, we find the following compact-form result:
\begin{equation}
    x^{-1}\frac{F^+_{4_1}(x,q)}{x^{1/2}-x^{-1/2}} = \sum_{d_1,\dots,d_6\geq 0}\frac{(-q^{1/2})^{\sum_{i,j}C_{ij}d_id_j}}{(q;q)_{d_i}}x_i^{d_i},
\end{equation}
where the quiver is
\begin{equation}
C=
\begin{bmatrix}
0 & 0 & 0 & 0 & 0 & 0 \\
0 & 0 & -1 & -1 & 0 & 0 \\
0 & -1 & 0 & 0 & 1 & 0 \\
0 & -1 & 0 & 1 & 1 & 0 \\
0 & 0 & 1 & 1 & 1 & 0 \\
0 & 0 & 0 & 0 & 0 & 1 \\
\end{bmatrix},
\quad
\boldsymbol{x} = 
\begin{bmatrix}
    x \\
    x \\
    x \\
    q^{1/2}x \\
    q^{1/2}x \\
    q^{1/2}x \\
\end{bmatrix}.
\end{equation}
One can then use the standard machinery of quiver equivalences from unlinking and permutohedra graphs (see Section \ref{sec:Permutohedra graphs}) to find other equivalent quivers for this example. Another interesting observation is that this quiver contains the unknot complement quiver which we found previously, but structurally it is very different from the HOMFLY-PT quivers (\ref{eq:figure-eight quivers}).


Note that knot complement quivers in general are not very well understood. There are some hints on their relation to geometry of knot complements, as discussed in \cite{Ekholm:2021irc} along with a few examples. Another consensus is that they should play a key role in understanding the categorification of quantum invariants of 3-manifolds.
There are simple families of knots, such as twist knots, for which $F_K$ invariants and their quivers can be computed by the methods described above. For example, positive double twist knots (such as $3_1$, $5_2$, $7_2$, etc.) can be extracted from the work of Lovejoy and Osburn on computation of colored Jones polynomials for double twist knots \cite{lovejoy2021colored}, and it was summarized in \cite{park2021inverted}. However, the general structure of the corresponding quivers remains mysterious. Identification of such structures for a larger classes of knots is an interesting research direction -- it must give additional insight on the role of quivers in low-dimensional topology.

\section*{Acknowledgements}

The authors would like to thank the organisers of workshop ``Knots, quivers and beyond'' which took place at IIT Bombay in February 2025, and in particular, Prof. Pichai Ramadevi for the warm hospitality.

\bibliography{refs}
\bibliographystyle{JHEP}
\end{document}

%% file: Figures/permut.tikz
\tikzset{every picture/.style={line width=0.75pt}} 

\begin{tikzpicture}[x=0.75pt,y=0.75pt,yscale=-0.5,xscale=0.5]

\draw    (1517.1,1036.79) .. controls (1737,1077.33) and (1814,1075.33) .. (2033.87,1036.79) ;
\draw    (1516.92,796.9) .. controls (1743,769.33) and (1800,763.33) .. (2033.69,796.9) ;
\draw    (1646.38,1031.43) -- (1904.59,1031.43) ;
\draw    (1646.56,802.46) -- (1904.77,802.46) ;
\draw   (1456.36,967.9) -- (1416.46,1037) -- (1336.67,1037) -- (1296.77,967.9) -- (1336.67,898.8) -- (1416.46,898.8) -- cycle ;
\draw    (1216.67,898) -- (1216.67,1037) ;
\draw  [fill={rgb, 255:red, 255; green, 255; blue, 255 }  ,fill opacity=1 ][line width=1.5]  (1211.17,1037) .. controls (1211.17,1033.96) and (1213.63,1031.5) .. (1216.67,1031.5) .. controls (1219.7,1031.5) and (1222.17,1033.96) .. (1222.17,1037) .. controls (1222.17,1040.04) and (1219.7,1042.5) .. (1216.67,1042.5) .. controls (1213.63,1042.5) and (1211.17,1040.04) .. (1211.17,1037) -- cycle ;
\draw  [fill={rgb, 255:red, 255; green, 255; blue, 255 }  ,fill opacity=1 ][line width=1.5]  (1211.17,898) .. controls (1211.17,894.96) and (1213.63,892.5) .. (1216.67,892.5) .. controls (1219.7,892.5) and (1222.17,894.96) .. (1222.17,898) .. controls (1222.17,901.04) and (1219.7,903.5) .. (1216.67,903.5) .. controls (1213.63,903.5) and (1211.17,901.04) .. (1211.17,898) -- cycle ;
\draw  [fill={rgb, 255:red, 255; green, 255; blue, 255 }  ,fill opacity=1 ][line width=1.5]  (1331.17,1037) .. controls (1331.17,1033.96) and (1333.63,1031.5) .. (1336.67,1031.5) .. controls (1339.7,1031.5) and (1342.17,1033.96) .. (1342.17,1037) .. controls (1342.17,1040.04) and (1339.7,1042.5) .. (1336.67,1042.5) .. controls (1333.63,1042.5) and (1331.17,1040.04) .. (1331.17,1037) -- cycle ;
\draw  [fill={rgb, 255:red, 255; green, 255; blue, 255 }  ,fill opacity=1 ][line width=1.5]  (1450.86,967.9) .. controls (1450.86,964.86) and (1453.32,962.4) .. (1456.36,962.4) .. controls (1459.39,962.4) and (1461.86,964.86) .. (1461.86,967.9) .. controls (1461.86,970.94) and (1459.39,973.4) .. (1456.36,973.4) .. controls (1453.32,973.4) and (1450.86,970.94) .. (1450.86,967.9) -- cycle ;
\draw  [fill={rgb, 255:red, 255; green, 255; blue, 255 }  ,fill opacity=1 ][line width=1.5]  (1410.96,898.8) .. controls (1410.96,895.76) and (1413.42,893.3) .. (1416.46,893.3) .. controls (1419.5,893.3) and (1421.96,895.76) .. (1421.96,898.8) .. controls (1421.96,901.83) and (1419.5,904.3) .. (1416.46,904.3) .. controls (1413.42,904.3) and (1410.96,901.83) .. (1410.96,898.8) -- cycle ;
\draw  [fill={rgb, 255:red, 255; green, 255; blue, 255 }  ,fill opacity=1 ][line width=1.5]  (1331.17,898.8) .. controls (1331.17,895.76) and (1333.63,893.3) .. (1336.67,893.3) .. controls (1339.7,893.3) and (1342.17,895.76) .. (1342.17,898.8) .. controls (1342.17,901.83) and (1339.7,904.3) .. (1336.67,904.3) .. controls (1333.63,904.3) and (1331.17,901.83) .. (1331.17,898.8) -- cycle ;
\draw  [fill={rgb, 255:red, 255; green, 255; blue, 255 }  ,fill opacity=1 ][line width=1.5]  (1291.27,967.9) .. controls (1291.27,964.86) and (1293.73,962.4) .. (1296.77,962.4) .. controls (1299.81,962.4) and (1302.27,964.86) .. (1302.27,967.9) .. controls (1302.27,970.94) and (1299.81,973.4) .. (1296.77,973.4) .. controls (1293.73,973.4) and (1291.27,970.94) .. (1291.27,967.9) -- cycle ;
\draw  [fill={rgb, 255:red, 255; green, 255; blue, 255 }  ,fill opacity=1 ][line width=1.5]  (1091.17,1037) .. controls (1091.17,1033.96) and (1093.63,1031.5) .. (1096.67,1031.5) .. controls (1099.7,1031.5) and (1102.17,1033.96) .. (1102.17,1037) .. controls (1102.17,1040.04) and (1099.7,1042.5) .. (1096.67,1042.5) .. controls (1093.63,1042.5) and (1091.17,1040.04) .. (1091.17,1037) -- cycle ;
\draw  [fill={rgb, 255:red, 255; green, 255; blue, 255 }  ,fill opacity=1 ][line width=1.5]  (1410.96,1037) .. controls (1410.96,1033.96) and (1413.42,1031.5) .. (1416.46,1031.5) .. controls (1419.5,1031.5) and (1421.96,1033.96) .. (1421.96,1037) .. controls (1421.96,1040.04) and (1419.5,1042.5) .. (1416.46,1042.5) .. controls (1413.42,1042.5) and (1410.96,1040.04) .. (1410.96,1037) -- cycle ;
\draw   (1666.56,996.69) -- (1597.46,956.79) -- (1597.46,877) -- (1666.56,837.1) -- (1735.66,877) -- (1735.66,956.79) -- cycle ;
\draw   (1884.77,996.69) -- (1815.66,956.79) -- (1815.66,877) -- (1884.77,837.1) -- (1953.87,877) -- (1953.87,956.79) -- cycle ;
\draw   (1735.66,877) -- (1815.66,877) -- (1815.66,957) -- (1735.66,957) -- cycle ;
\draw   (1577.46,842.36) -- (1646.74,802.36) -- (1666.74,837) -- (1597.46,877) -- cycle ;
\draw   (1884.59,996.79) -- (1953.87,956.79) -- (1973.87,991.43) -- (1904.59,1031.43) -- cycle ;
\draw   (1904.77,802.46) -- (1974.05,842.46) -- (1954.05,877.1) -- (1884.77,837.1) -- cycle ;
\draw   (1597.1,956.79) -- (1666.38,996.79) -- (1646.38,1031.43) -- (1577.1,991.43) -- cycle ;
\draw    (1517.46,797) -- (1577.46,842.36) ;
\draw    (1973.87,991.43) -- (2033.87,1036.79) ;
\draw    (1517.46,1037) -- (1577.1,991.43) ;
\draw    (1974.05,842.46) -- (2033.69,796.9) ;
\draw    (1517.46,797) -- (1517.46,1037) ;
\draw    (2033.87,796.79) -- (2033.87,1036.79) ;
\draw  [fill={rgb, 255:red, 255; green, 255; blue, 255 }  ,fill opacity=1 ][line width=1.5]  (1511.6,796.79) .. controls (1511.6,793.75) and (1514.06,791.29) .. (1517.1,791.29) .. controls (1520.14,791.29) and (1522.6,793.75) .. (1522.6,796.79) .. controls (1522.6,799.83) and (1520.14,802.29) .. (1517.1,802.29) .. controls (1514.06,802.29) and (1511.6,799.83) .. (1511.6,796.79) -- cycle ;
\draw  [fill={rgb, 255:red, 255; green, 255; blue, 255 }  ,fill opacity=1 ][line width=1.5]  (1571.96,842.36) .. controls (1571.96,839.32) and (1574.42,836.86) .. (1577.46,836.86) .. controls (1580.5,836.86) and (1582.96,839.32) .. (1582.96,842.36) .. controls (1582.96,845.4) and (1580.5,847.86) .. (1577.46,847.86) .. controls (1574.42,847.86) and (1571.96,845.4) .. (1571.96,842.36) -- cycle ;
\draw  [fill={rgb, 255:red, 255; green, 255; blue, 255 }  ,fill opacity=1 ][line width=1.5]  (1641.24,802.36) .. controls (1641.24,799.32) and (1643.7,796.86) .. (1646.74,796.86) .. controls (1649.78,796.86) and (1652.24,799.32) .. (1652.24,802.36) .. controls (1652.24,805.4) and (1649.78,807.86) .. (1646.74,807.86) .. controls (1643.7,807.86) and (1641.24,805.4) .. (1641.24,802.36) -- cycle ;
\draw  [fill={rgb, 255:red, 255; green, 255; blue, 255 }  ,fill opacity=1 ][line width=1.5]  (1661.06,837.1) .. controls (1661.06,834.07) and (1663.52,831.6) .. (1666.56,831.6) .. controls (1669.6,831.6) and (1672.06,834.07) .. (1672.06,837.1) .. controls (1672.06,840.14) and (1669.6,842.6) .. (1666.56,842.6) .. controls (1663.52,842.6) and (1661.06,840.14) .. (1661.06,837.1) -- cycle ;
\draw  [fill={rgb, 255:red, 255; green, 255; blue, 255 }  ,fill opacity=1 ][line width=1.5]  (1591.6,876.79) .. controls (1591.6,873.75) and (1594.06,871.29) .. (1597.1,871.29) .. controls (1600.14,871.29) and (1602.6,873.75) .. (1602.6,876.79) .. controls (1602.6,879.83) and (1600.14,882.29) .. (1597.1,882.29) .. controls (1594.06,882.29) and (1591.6,879.83) .. (1591.6,876.79) -- cycle ;
\draw  [fill={rgb, 255:red, 255; green, 255; blue, 255 }  ,fill opacity=1 ][line width=1.5]  (1591.6,956.79) .. controls (1591.6,953.75) and (1594.06,951.29) .. (1597.1,951.29) .. controls (1600.14,951.29) and (1602.6,953.75) .. (1602.6,956.79) .. controls (1602.6,959.83) and (1600.14,962.29) .. (1597.1,962.29) .. controls (1594.06,962.29) and (1591.6,959.83) .. (1591.6,956.79) -- cycle ;
\draw  [fill={rgb, 255:red, 255; green, 255; blue, 255 }  ,fill opacity=1 ][line width=1.5]  (1661.06,996.69) .. controls (1661.06,993.65) and (1663.52,991.19) .. (1666.56,991.19) .. controls (1669.6,991.19) and (1672.06,993.65) .. (1672.06,996.69) .. controls (1672.06,999.73) and (1669.6,1002.19) .. (1666.56,1002.19) .. controls (1663.52,1002.19) and (1661.06,999.73) .. (1661.06,996.69) -- cycle ;
\draw  [fill={rgb, 255:red, 255; green, 255; blue, 255 }  ,fill opacity=1 ][line width=1.5]  (1640.88,1031.43) .. controls (1640.88,1028.4) and (1643.34,1025.93) .. (1646.38,1025.93) .. controls (1649.42,1025.93) and (1651.88,1028.4) .. (1651.88,1031.43) .. controls (1651.88,1034.47) and (1649.42,1036.93) .. (1646.38,1036.93) .. controls (1643.34,1036.93) and (1640.88,1034.47) .. (1640.88,1031.43) -- cycle ;
\draw  [fill={rgb, 255:red, 255; green, 255; blue, 255 }  ,fill opacity=1 ][line width=1.5]  (1571.6,991.43) .. controls (1571.6,988.4) and (1574.06,985.93) .. (1577.1,985.93) .. controls (1580.14,985.93) and (1582.6,988.4) .. (1582.6,991.43) .. controls (1582.6,994.47) and (1580.14,996.93) .. (1577.1,996.93) .. controls (1574.06,996.93) and (1571.6,994.47) .. (1571.6,991.43) -- cycle ;
\draw  [fill={rgb, 255:red, 255; green, 255; blue, 255 }  ,fill opacity=1 ][line width=1.5]  (1511.6,1036.79) .. controls (1511.6,1033.75) and (1514.06,1031.29) .. (1517.1,1031.29) .. controls (1520.14,1031.29) and (1522.6,1033.75) .. (1522.6,1036.79) .. controls (1522.6,1039.83) and (1520.14,1042.29) .. (1517.1,1042.29) .. controls (1514.06,1042.29) and (1511.6,1039.83) .. (1511.6,1036.79) -- cycle ;
\draw  [fill={rgb, 255:red, 255; green, 255; blue, 255 }  ,fill opacity=1 ][line width=1.5]  (1730.16,956.79) .. controls (1730.16,953.75) and (1732.63,951.29) .. (1735.66,951.29) .. controls (1738.7,951.29) and (1741.16,953.75) .. (1741.16,956.79) .. controls (1741.16,959.83) and (1738.7,962.29) .. (1735.66,962.29) .. controls (1732.63,962.29) and (1730.16,959.83) .. (1730.16,956.79) -- cycle ;
\draw  [fill={rgb, 255:red, 255; green, 255; blue, 255 }  ,fill opacity=1 ][line width=1.5]  (1730.16,877) .. controls (1730.16,873.96) and (1732.63,871.5) .. (1735.66,871.5) .. controls (1738.7,871.5) and (1741.16,873.96) .. (1741.16,877) .. controls (1741.16,880.04) and (1738.7,882.5) .. (1735.66,882.5) .. controls (1732.63,882.5) and (1730.16,880.04) .. (1730.16,877) -- cycle ;
\draw  [fill={rgb, 255:red, 255; green, 255; blue, 255 }  ,fill opacity=1 ][line width=1.5]  (1810.16,877) .. controls (1810.16,873.96) and (1812.63,871.5) .. (1815.66,871.5) .. controls (1818.7,871.5) and (1821.16,873.96) .. (1821.16,877) .. controls (1821.16,880.04) and (1818.7,882.5) .. (1815.66,882.5) .. controls (1812.63,882.5) and (1810.16,880.04) .. (1810.16,877) -- cycle ;
\draw  [fill={rgb, 255:red, 255; green, 255; blue, 255 }  ,fill opacity=1 ][line width=1.5]  (1810.16,957) .. controls (1810.16,953.96) and (1812.63,951.5) .. (1815.66,951.5) .. controls (1818.7,951.5) and (1821.16,953.96) .. (1821.16,957) .. controls (1821.16,960.04) and (1818.7,962.5) .. (1815.66,962.5) .. controls (1812.63,962.5) and (1810.16,960.04) .. (1810.16,957) -- cycle ;
\draw  [fill={rgb, 255:red, 255; green, 255; blue, 255 }  ,fill opacity=1 ][line width=1.5]  (1879.09,996.79) .. controls (1879.09,993.75) and (1881.55,991.29) .. (1884.59,991.29) .. controls (1887.62,991.29) and (1890.09,993.75) .. (1890.09,996.79) .. controls (1890.09,999.83) and (1887.62,1002.29) .. (1884.59,1002.29) .. controls (1881.55,1002.29) and (1879.09,999.83) .. (1879.09,996.79) -- cycle ;
\draw  [fill={rgb, 255:red, 255; green, 255; blue, 255 }  ,fill opacity=1 ][line width=1.5]  (1899.09,1031.43) .. controls (1899.09,1028.4) and (1901.55,1025.93) .. (1904.59,1025.93) .. controls (1907.62,1025.93) and (1910.09,1028.4) .. (1910.09,1031.43) .. controls (1910.09,1034.47) and (1907.62,1036.93) .. (1904.59,1036.93) .. controls (1901.55,1036.93) and (1899.09,1034.47) .. (1899.09,1031.43) -- cycle ;
\draw  [fill={rgb, 255:red, 255; green, 255; blue, 255 }  ,fill opacity=1 ][line width=1.5]  (1968.37,991.43) .. controls (1968.37,988.4) and (1970.83,985.93) .. (1973.87,985.93) .. controls (1976.91,985.93) and (1979.37,988.4) .. (1979.37,991.43) .. controls (1979.37,994.47) and (1976.91,996.93) .. (1973.87,996.93) .. controls (1970.83,996.93) and (1968.37,994.47) .. (1968.37,991.43) -- cycle ;
\draw  [fill={rgb, 255:red, 255; green, 255; blue, 255 }  ,fill opacity=1 ][line width=1.5]  (1951.6,956.79) .. controls (1951.6,953.75) and (1954.06,951.29) .. (1957.1,951.29) .. controls (1960.14,951.29) and (1962.6,953.75) .. (1962.6,956.79) .. controls (1962.6,959.83) and (1960.14,962.29) .. (1957.1,962.29) .. controls (1954.06,962.29) and (1951.6,959.83) .. (1951.6,956.79) -- cycle ;
\draw  [fill={rgb, 255:red, 255; green, 255; blue, 255 }  ,fill opacity=1 ][line width=1.5]  (1948.55,877.1) .. controls (1948.55,874.07) and (1951.01,871.6) .. (1954.05,871.6) .. controls (1957.08,871.6) and (1959.55,874.07) .. (1959.55,877.1) .. controls (1959.55,880.14) and (1957.08,882.6) .. (1954.05,882.6) .. controls (1951.01,882.6) and (1948.55,880.14) .. (1948.55,877.1) -- cycle ;
\draw  [fill={rgb, 255:red, 255; green, 255; blue, 255 }  ,fill opacity=1 ][line width=1.5]  (1968.55,842.46) .. controls (1968.55,839.43) and (1971.01,836.96) .. (1974.05,836.96) .. controls (1977.08,836.96) and (1979.55,839.43) .. (1979.55,842.46) .. controls (1979.55,845.5) and (1977.08,847.96) .. (1974.05,847.96) .. controls (1971.01,847.96) and (1968.55,845.5) .. (1968.55,842.46) -- cycle ;
\draw  [fill={rgb, 255:red, 255; green, 255; blue, 255 }  ,fill opacity=1 ][line width=1.5]  (1899.27,802.46) .. controls (1899.27,799.43) and (1901.73,796.96) .. (1904.77,796.96) .. controls (1907.8,796.96) and (1910.27,799.43) .. (1910.27,802.46) .. controls (1910.27,805.5) and (1907.8,807.96) .. (1904.77,807.96) .. controls (1901.73,807.96) and (1899.27,805.5) .. (1899.27,802.46) -- cycle ;
\draw  [fill={rgb, 255:red, 255; green, 255; blue, 255 }  ,fill opacity=1 ][line width=1.5]  (1879.27,837.1) .. controls (1879.27,834.07) and (1881.73,831.6) .. (1884.77,831.6) .. controls (1887.8,831.6) and (1890.27,834.07) .. (1890.27,837.1) .. controls (1890.27,840.14) and (1887.8,842.6) .. (1884.77,842.6) .. controls (1881.73,842.6) and (1879.27,840.14) .. (1879.27,837.1) -- cycle ;
\draw  [fill={rgb, 255:red, 255; green, 255; blue, 255 }  ,fill opacity=1 ][line width=1.5]  (2028.19,796.9) .. controls (2028.19,793.86) and (2030.65,791.4) .. (2033.69,791.4) .. controls (2036.73,791.4) and (2039.19,793.86) .. (2039.19,796.9) .. controls (2039.19,799.93) and (2036.73,802.4) .. (2033.69,802.4) .. controls (2030.65,802.4) and (2028.19,799.93) .. (2028.19,796.9) -- cycle ;
\draw  [fill={rgb, 255:red, 255; green, 255; blue, 255 }  ,fill opacity=1 ][line width=1.5]  (2028.37,1036.79) .. controls (2028.37,1033.75) and (2030.83,1031.29) .. (2033.87,1031.29) .. controls (2036.91,1031.29) and (2039.37,1033.75) .. (2039.37,1036.79) .. controls (2039.37,1039.83) and (2036.91,1042.29) .. (2033.87,1042.29) .. controls (2030.83,1042.29) and (2028.37,1039.83) .. (2028.37,1036.79) -- cycle ;

\draw (1096.67,1072) node [anchor=north] [inner sep=0.75pt]    {$\Pi_{1}$};
\draw (1216.67,1072) node [anchor=north] [inner sep=0.75pt]    {$\Pi_{2}$};
\draw (1376.67,1072) node [anchor=north] [inner sep=0.75pt]    {$\Pi_{3}$};
\draw (1768.67,1072) node [anchor=north] [inner sep=0.75pt]    {$\Pi_{4}$};

\end{tikzpicture}